# Statistical advances and challenges for analyzing correlated high dimensional SNP data in genomic study for complex diseases*

## Yulan Liang

*Department of Biostatistics University at Buffalo, The State University of New York, Buffalo, NY 14214, USA*
*e-mail:* yliang@buffalo.edu

## Arpad Kelemen

*Department of Neurology, Buffalo Neuroimaging Analysis Center, The Jacobs Neurological Institute, University at Buffalo, The State University of New York, 100 High Street, Buffalo, NY 14203, USA*
*e-mail:* akelemen@buffalo.edu

**Abstract:** Recent advances of information technology in biomedical sciences and other applied areas have created numerous large diverse data sets with a high dimensional feature space, which provide us a tremendous amount of information and new opportunities for improving the quality of human life. Meanwhile, great challenges are also created driven by the continuous arrival of new data that requires researchers to convert these raw data into scientific knowledge in order to benefit from it. Association studies of complex diseases using SNP data have become more and more popular in biomedical research in recent years. In this paper, we present a review of recent statistical advances and challenges for analyzing correlated high dimensional SNP data in genomic association studies for complex diseases. The review includes both general feature reduction approaches for high dimensional correlated data and more specific approaches for SNPs data, which include unsupervised haplotype mapping, tag SNP selection, and supervised SNPs selection using statistical testing/scoring, statistical modeling and machine learning methods with an emphasis on how to identify interacting loci.

Received June 2007.

## Contents

---

*This paper was accepted by Michael Kosorok, Associate Editor for the IMS.

## 1. Introduction

Correlating genetic variations in DNA sequences with phenotypic differences has been one of the grand challenges in biomedical research. Substantial efforts have been made to obtain all common genetic variations in humans, including single nucleotide polymorphisms (SNPs), deletions and insertions [13]. SNPs are single base pair positions in genomic DNA at which different sequence alternatives (alleles) exist in normal individuals in some population(s), wherein the least frequent allele has an abundance of 1% or greater [13]. In practice, the term "SNP" is used more loosely. Restricting the attention to common SNPs with minor allele frequency bigger than a certain cutoff, e.g. 1% will help to filter out some "recent" mutations. SNPs are believed to alter the risk for developing particular diseases. It is, however, very unlikely that individual SNPs play an important role in the development of complex diseases. Instead, high-order interactions of SNPs are supposed to explain the differences between low and high risk population groups.

The HapMap Project has collected genotypes of millions of SNPs from populations with ancestry from Africa, Asia and Europe and makes this information freely available in the public domain [93–95]. To find evidence of association in this huge data set is a grand challenge now. Therefore, there is a great need, conceptually as well as computationally, to develop advanced robust algorithms and analytical methods for characterizing genetic variations that are non-redundant and identify the target SNPs that are most likely to affect the phenotypes and ultimately contribute to disease development.

Exploiting information redundancy due to associations between SNP markers potentially reduces the efforts in terms of time and cost for genetic association studies [75]. However, the efficacy of searching for an optimal set of SNPs has not been as successful as expected in theory. One primary cause is the high dimensionality with highly correlated features/SNPs that can hinder the power of the identification of small to moderate genetic effects in complex diseases. The need to incorporate covariates of other environmental risk factors as effect modifiers or confounders further worsens the "curse of dimensionality" problem in mapping genes for complex diseases [16]. Therefore, feature selection for massive genomic data in high dimensions has become one of the main tasks to be tackled with statistical and computational efforts in the past decade.

## 2. Feature selection methods for high dimensional problems

The computational and statistical methods that address the "curse of dimensionality" problem in genomic research can be grouped into three categories: filtering, wrapper, and embedded methods. Filtering methods select feature subsets independently from the learning classifiers and do not incorporate learning. Therefore, filtering methods are fast [10; 60; 69; 109]. A weakness of filtering methods is that they only consider the individual features in isolation and ignore the possible interaction among them. Yet, the combination of these features may have a combined effect that does not necessarily follow from the individual performances of features in the group [73]. One of the consequences of filtering methods is that we may end up with many highly correlated features/SNPs with highly redundant information that worsens the classification and prediction performance. If there is a limit on the number of features to be chosen, then we may not be able to include all the informative ones.

To address this problem in filtering methods, wrapper methods wrap around a particular learning algorithm that can assess the selected feature subsets in terms of the estimated classification errors and then build the final classifier [44]. Wrapper methods use a learning machine to measure the quality of subsets of features. One of the well-known wrapper methods for feature selection is Support Vector Machine Recursive Feature Elimination, which refines the optimum feature set by using a Support Vector Machine, [33]. The idea of SVMRFE is that the orientation of the separating hyper-plane found by the SVM can be used to select informative features: if the plane is orthogonal to a particular feature dimension, then that feature is informative, and vice versa. SVMRFE uses the weights of a SVM classifier to produce a feature ranking, and then eliminates the feature with smallest weight magnitude recursively.

Wrapper methods can be used with arbitrary classifiers and can notably reduce the number of features and significantly improve the classification accuracy [63; 79]. However, wrapper methods have the drawback that they do not incorporate knowledge about the specific structure of the classification or regression function [52]. Moreover, they are more computationally expensive since they need to evaluate a cross-validation scheme at each iteration.

With much better computational efficiency and similar performance to wrapper methods, a relatively new class of approaches for feature selection called "embedded methods" has become available in the literature. Lal et al. [52] provide the detailed mathematical formulations of embedded methods. Embedded methods process feature selection simultaneously with the learning classifier and the feature selection can not be separated from the learning. For example, Weston et al. [107] measure the importance of a feature using a bound that is valid for Support Vector Machines only, thus it is not possible to use this method with, for example, decision trees.

Therefore the structure of the class of functions under consideration plays a crucial role. For an embedded method, every subset of features is modeled by a vector $\sigma \in \{0,1\}^n$ of indicator variables, $\sigma_i := 0$ indicating that a feature is present in a subset and $\sigma_i := 1$ indicating that a feature is absent (i = 1,...,n).

A parameterized family of classification or regression functions are given as follows: $f : \Lambda \times \Re^n \to \Re, (\alpha, x) \propto f(\alpha, x)$. The goal of an embedded method is to find a vector of indicator variables $\sigma^* \in \{0, 1\}^n$ and $\alpha^* \in \Lambda$ that minimize the expected risk $R(\alpha, \sigma) = \int L[f(\alpha, \sigma * x), y] dF(x, y)$, where $*$ denotes the pointwise product, $L$ is a loss function and $P$ is a measure on the domain of the training data $(X; Y)$.

One may impose some additional constraints for penalty or regularizations to achieve sparseness: $s(\sigma) \le \sigma_0$, where $s : [0, 1]^n \to \Re^+$ measures the sparsity of a given indicator vector $\sigma$. For example, $s$ could be defined as: $s(\sigma) := l_0(\sigma) \le \sigma_0$, that is to bound the zero "norm" $l_0(\sigma)$, which counts the number of nonzero entries in $\sigma$. The L1-norm, L2-norm, and L$\infty$-norm or the elastic-net penalty, a mixture of the L2-norm and the L1-norm penalties [105] are also proposed to achieve automatic feature selections by shrinking the fitted coefficients toward zero. These automatic feature selection methods also benefit from the reduction in the fitted coefficients' variance.

One of the merits of an embedded method is that it intends to find the feature subset of a certain size that leads to the best possible generalization or equivalently to minimal risk, which can be seen from the above formulation. Therefore, the function that measures the quality of a scaling factor can be evaluated faster than a cross-validation error estimation procedure. Moreover, they turn the multiple testing problems for feature selection into an optimization problem in the nonparametric setting. Some recent studies [90; 105] have shown that they are more computational efficient and asymptotically optimal for high dimensional data.

Embedded methods tend to have higher capacity than filtering methods and are therefore more likely to overfit. We thus expect filtering methods to perform better if only a small amount of training data is available. Embedded methods eventually outperform filtering methods as the number of training samples increase. LASSO proposed by Tibshirani [97; 98], logic regression with the regularized Laplacian prior [51] and Bayesian regularized neural network with automatic relevance determination [56] are examples of embedded methods.

Note that the three feature reduction methods, filter, wrapper and embedded methods discussed in this section may perform differently when applied to categorical SNP data instead of continuous gene expression data, in which there are only three genotypes, two homozygous genotypes and one heterozygous genotype. Next we will focus on the review of the recently developed categorical SNP data reduction methods in genome wide association studies.

## 3. SNP selections in genome-wide association studies

A major aim of association studies is the identification of polymorphisms, usually single nucleotide polymorphisms (SNPs) associated with a trait or disease status. There are several major computational and statistical tracks for SNP selections, which we will review next [3; 18; 25; 34; 35; 40; 46; 48; 59; 115].

### 3.1. Statistical measures and testing for SNP-disease association

Specifically, in genome-wide disease association studies, various statistical measures and testing based approaches have been proposed for selecting a subset of SNPs [17; 30; 36; 57; 85; 89]. These include Linkage Disequilibrium (LD) based SNP selection and supervised SNP selection. Linkage Disequilibrium based methods for selecting a maximally informative set of SNPs for association analyses were developed first [24; 92; 101; 102; 108]. For instance, Zhang and Jin [114] identified tagSNPs from haplotype data in two steps; first, they identified haplotype blocks and then identified tagSNPs that best distinguish the haplotypes within a haplotype block. This method is applicable for all types of association studies. Anderson and Novembre [1] and Mannila et al. [61] proposed finding haplotype block boundaries using minimum description length. Entropy-based measure for SNP selections were proposed by Hampe, Schreiber, and Krawczak [36] and Zhao, Boerwinkle, and Xiong [117]. Beckmann et al. [8] presented Mantel statistics for SNP selections and disease mapping purposes by using haplotype sharing to correlate temporal and spatial distributions of cancer in a generalized regression model.

A sliding window approach developed by Neale and Sham [68] combines p-values from multiple independent tests using $\chi^2 = -2\sum_{i=1}^{m} log(p_i) \sim \chi^2_{2m}$. Here, $p_i$ is the p-value of association between $SNP_i$ and presence of disease, and $m$ is the number of SNPs in the sliding window. The test statistic $\chi^2$ has a chi-square distribution with $2m$ degrees of freedom. The sliding window incorporates the ordering of SNPs on the chromosome and merges results across adjacent windows to detect chromosome regions with significant associations [27; 84; 113]. However, it does not consider the distance between them and the implicit assumption is that the SNPs are equally spaced.

The scan statistic [26; 39; 54; 91; 104] does account for the spacing and ordering of SNPs on the chromosome, but it does not consider gene-gene interactions. For instance, Sun, et al. [91] developed a chromosomal scan statistic approach, which includes two parts: (i) Identifying SNP clusters; (ii) Identifying SNP clusters with significant disease association. This scan method assumes the position of each SNP is randomly determined by a Poisson process. The lengths between two adjacent SNPs have an exponential distribution and the sum-of-lengths between SNPs has a Gamma distribution. Under the above assumptions, the clusters of SNPs are first identified by testing the hypothesis that whether the observed length between a set of SNPs (combined interval between these SNPs) is equal (null hypothesis) or less than (alternative hypothesis) the expected length. Rejection of the null hypothesis identifies this group of SNPs as cluster. To further identify SNP clusters with significant disease association, disease outcomes are incorporated and Pearson Chi-square p-values are computed for associations of significance.

Other test statistic approaches, such as the score statistic [81; 82], and weighted-average statistic [87] for disease mapping in case-control studies were also proposed for SNP selection in genetic association studies. Cheng et al. [20] propose using the expectation maximization (EM) algorithm to estimate haplotype fre-quencies of multiple linked SNPs, and follow this by constructing a contingency table statistic S for LD analysis based on the estimated haplotype frequencies. An empirical p-value is obtained based on the null distribution of the maximum of S (S*) from a large number (e.g., 1,000 or more) of randomized permutations. This method is developed for mapping functional sites or regions from case-control data using haplotypes of multiple linked SNPs.

All these conventional test based filter approaches estimate the association between each SNP (or multiple SNPs) and a phenotype, and then use the corresponding p-values to prioritize the results. One drawback is that one may end up with many highly correlated SNPs or genes with high redundancy information, which can be hurdles for further classifications and predictions. Also the test based approaches can not incorporate many environmental factors to account for gene-environment interactions. Furthermore, the non-independence of SNPs in physical proximity (Linkage Disequilibrium) may cause problems for multiple testing scenarios with correlated tests [6; 7; 23; 70; 80; 112]. Simple corrections may lead to either conservative p-values if Bonferroni correction is used or become computationally expensive, if permutation is used [84]. Nyholt [70] proposed a method for efficiently accounting for multiple testing of many SNPs in an association study that involves estimating an "effective number" of independent tests, and then adjusting the smallest observed p-value using Sidak's formula based on this number of tests. Salyakina et al. [80] further evaluated this method.

Note that the "multiple testing problem" discussed here differs from the "curse of dimensionality problem", so it poses different challenges. "Multiple testing problem" is caused by the high dimensionality of the predictors (including features plus possible interactions of features) and the complex correlation structures of the predictors, while the "curse of dimensionality problem" arises when considering the interaction of many features, i.e., there are not enough observations in each combination of those features.

Last, but not least, these existing testing based approaches ignore some information about the SNPs, such as sub-structures of the underlying population (admixture problem). This may lead to spurious results as well as suffer from low power. This may explain why reproducibility has become a major issue in genomic association studies for complex diseases. The same data set can show a highly significant association with one method, whereas a different method shows no or only a marginal association. Also, given the low prior probability of causality for each SNP in the genome, rigorous standards of statistical significance are needed for genome-wide association studies in order to avoid a flood of false-positive results. Multiple replications in large samples may provide the most straightforward path in identifying robust and broadly relevant associations.

### 3.2. Supervised statistical models and statistical learning algorithms

In order to incorporate environmental factors and other covariates/confounders into the genomic association studies, various model based approaches have been developed. He and Zelikovsky [37] proposed tagSNPs for unphased genotypes based on multiple linear regressions. Durrant et al. [28] adopted a logistic-regression model applicable to whole-genome screens using sliding windows; it controls for other (continuous) confounders and gene-environment interactions. Yet, they have to make assumptions on the disease model, which is usually unknown in practice. Moreover, the effects of violations of these assumptions are unpredictable in general. Baker [5] applied a simple loglinear model for haplotype effects in a case-control study involving two unphased genotypes.

The haplotype trend regression, developed by Zaykin et al. [111], fits a model of additive effects of haplotypes that takes a series of marker genotypes, computes haplotype probabilities for each observation using the composite haplotype method (CHM), and forms a linear regression on the response using the haplotype probabilities as the regression matrix. A nonparametric method called Haplotype Pattern Mining (HPM) was proposed to identify disease associated haplotype patterns from case-control data. HPM has two steps: In step I, given the data-markers, haplotypes, and phenotypes, the goal is to output all haplotype patterns that are strongly associated with the disease status for a given value of the association threshold; In step II, it is to find the "gene location", by counting frequency that one marker appears in the haplotype patterns identified in the first step. Since the HPM method utilizes the disease status (case/control), it is a supervised mining approach. Toivonen et al. [99] showed that HPM does not require any assumptions on the inheritance patterns and has good localization power, even when the number of phenocopies is large. Knorr-Held and Rue [49] developed Markov random field models on block updating for disease mapping. Other model-based approaches that can take into account the spatial correlation between markers were also proposed [14; 31; 32; 42; 96; 100; 103; 106].

Recently, Schwender and Ickstadt [83] demonstrated logic regression based identification of SNP interactions for the disease status in case-control study and proposed two measures for quantifying the importance of feature interactions for classification. In comparison with some well-known classification methods such as CART [12], Random Forests [11], and other regression procedures [108], logic regression has shown a good classification performance when applied to SNP data. When fitting with categorical features/variables in the model based approaches, i.e. the genotype measurements with two homozygous genotypes and one heterozygous genotype, we often define a set of dummy variables that represent a single categorical feature/variable. In order to select the set/group of dummy variables that represent a single categorical feature/variable/SNP simultaneously, Yuan and Lin [110] proposed the group-Lars and the group-Lasso methods. Park and Hastie [72] proposed several regularization path algorithms with grouped feature/variable selection for modeling gene-gene interactions. Multifactor dimensionality reduction has been proposed and implemented for SNPs data reduction by Coffey et al. [22], Ritchie et al. [77] and Moore et al. [64].

### 3.3. Unsupervised haplotype mapping approaches

Haplotype density based clustering algorithms and clustering techniques based on the degree of haplotype sharing in affected individuals for haplotype mapping were developed recently. These approaches have advantages of robustness since they are nonparametric and require fewer assumptions in modeling. Fu et al. [29] and Zhang et al. [116] proposed Bayesian models for the analysis of genetic structure when populations are correlated. Liu et al. [58] employed a Bayesian approach to model positions of the historical recombinations and mutation events that produced the observed haplotypes from an initial set of founders by accounting for all sources of uncertainties. They employed Monte Carlo Markov Chain (MCMC) method for parameter estimation and assigned haplotypes to clusters representing allele heterogeneity. Molitor et al. [62] modeled haplotype risks using clusters obtained from a probability model, but their method does not take phenocopies into consideration. Both methods were developed mainly for haplotype fine mapping and do not scale up for whole-genome screens very well.

Other algorithms for SNPs are hierarchical clustering and graph methods [2; 55]. Principal Component Analysis with multiple genotype frequencies was also applied to select a subset of correlated SNPs that capture multiple genotype variability in the region [9; 57]. However, whether Principal Component Analysis is a suitable tool for categorical SNPs information is arguable, since it is more appropriate for continuous scale data. The related correspondence analysis may be more suitable, but the interpretation of the results from correspondence analysis reveals many challenges.

### 3.4. Computational intelligence approaches

Computational intelligence systems [47; 74] hold a great promise for tackling the tasks and challenges posed by large, diverse, genomic data for complex diseases. Some of these challenges are the identification of gene-gene and gene-environment interactions [4; 43; 50; 66; 78], dealing with the notorious "curse of dimensionality", the uncertainty, and unclear, fuzzy boundaries of phenotypes for complex diseases [76; 88]. Techniques include neural networks [71], genetic algorithms [21], genetic programming [65], evolutionary trees [53], evolutionary algorithms [41] and various hybrid approaches. For instance, Moore [64] developed a hybrid genetic programming (GP) with a multifactor dimensionality reduction method to pick SNPs for epistasis.

Motsinger, et al. [67] applied a genetic programming neural network (GPNN) approach for detecting epistasis in case-control studies for SNPs data. They evaluated the power of GPNN for identifying high-order gene-gene interactions and applied GPNN to a real data on Parkinson's disease. They developed a Grammatical Evolution Neural Network (GENN), a machine-learning approach to detect gene-gene and gene-environment interactions in high dimensional genetic epidemiological data. Furthermore, they proposed an Ensemble Learning Approach for Set association (ELAS) to detect a set of interacting loci that predicts the complex trait. An important advantage of the hybrid approach is that any form of expert knowledge could be used to guide the stochastic search algorithm to identify epistatic SNPs in the absence of marginal effects.

## 4. Other challenges in genetic association studies of complex diseases

An important challenge that faces molecular association studies in the post genomic era is to understand the interconnections from a network of genes and their products that are modified by a variety of environmental factors [15; 45]. The variety of phenotype definitions leads to a multiplicity of tests that involve a large number of comparisons that often result in less power. The need for adequate algorithms and models for reducing biological and statistical redundancy from thousands of SNPs and finding an optimal set of SNPs associated with diseases are pressing for common complex diseases. Dealing with many dependent association tests is one of the emerging issues on the statistical/computational side.

For SNP-disease data, in addition to being large, redundant, diverse and distributed, three important characteristics pose challenges for data analysis and modeling: (1) heterogeneity, (2) a constantly evolving biological nature and (3) complexity. Firstly, there is the heterogeneity of SNP data, in the sense that i) the population data involves the population substructure or admixture problem and there is locus heterogeneity where a large fraction of the prevalence is due to phenocopies; and ii) there is a wide array of data types, including categorical, continuous, sequence data, as well as temporal, incomplete and missing data. Such data sets are large with a lot of redundancy in SNP and haplotype databases. Secondly, they are very dynamic and continuously evolving, which means that special knowledge is required when designing the modeling techniques. Lastly, but most importantly, these SNP and haplotype data are complex with intrinsic features and subtle patterns, in the sense that they are very rich in associated complex phenotype traits.

The difficulty in a SNPs association study is increased by the nature of complex diseases [38]. Typically, the contribution of single genes as well as of single environmental risk factors is small to moderate. Furthermore, most complex diseases result from gene-gene and gene-environment interactions [19]. By disregarding interactions, relative risks of individual genetic variants are expected to be small. Disregarding gene-environment interaction also weakens exposure-disease and gene-disease associations. In complex diseases, it is likely that a combination of genes predisposes for the disease and environmental factors aggravate the impact of these genes and therefore are jointly responsible for disease development in populations (known as epistatic effect). In addition, environmental factors, which seem to have only a moderate impact at the population level might have larger relative risks in subpopulations with certain genetic predispositions. There are major methodological challenges in the study of gene-gene and gene-environment interactions.

Other open questions and challenges for new computational approaches in analyzing the associations between genetic markers, such as SNPs in complex diseases involve several hierarchical levels. First level of complexity: How to analyze multiple SNPs in a single gene? How to analyze interactions among multiple SNPs in a single gene? Second level of complexity: How to analyze multiple SNPs in multiple genes? How to analyze interactions among multiple SNPs in multiple genes? Third level of complexity: How to analyze interactions among multiple SNPs in multiple genes and environmental factors? Fourth level of complexity: How to analyze associations between SNPs in single or multiple genes and quantitative traits? How to identify and quantify the percentage of the association between genes and diseases explained by the association between the same gene and quantitative traits, taking into consideration single genes, multiple genes and environmental factors. Lastly, the ultimate goal in genetic/genomic analysis is to build direct or indirect causal association between genetic variants and phenotypes/disease status, but the difficulty here is that we do not know if there is association between the SNPs and the disease. However, with the development of computational/statistical approaches, we may be able to identify these causal associations and construct the pathways related to complex diseases.

## 5. Discussion

New advances in human genome research have drawn tremendous attention of researchers from multiple fields, including both theoretical scientists and applied researchers, especially in the statistical field. Huge amounts of continuously growing large-scale genomic, proteomic and clinical data for complex diseases and phenotype traits have posed ever greater challenges for the computational field. Multiple whole genome wide association studies have already been completed and have resulted in novel and promising genetic variants for various diseases. In this paper we presented a survey of recent advances and some promises of designing, developing and implementing statistical/computational methods for identifying SNP markers responsible for common, complex, chronic diseases, such as diabetes, cancer, multiple sclerosis, and cardiovascular disease and for tackling the challenges, such as gene-gene and gene-environment interactions along with the notorious "curse of dimensionality" problem. Success in identifying SNPs and haplotypes conferring susceptibility or resistance to common diseases will provide a deeper understanding of the architecture of the disease, the risks, and offer a more powerful diagnostic tool and predictive treatment.

## References

[1] ANDERSON, E.C. and NOVEMBRE, J. (2003). Finding haplotype block boundaries by using the minimum-description-length principle. *American Journal of Human Genetics* **73** 336–354.

[2] Ao, S., Yip, K., Ng, M., Cheung, D., Fong, P.Y., Melhado, I. and Sham, P.C. (2005). CLUSTAG: hierarchical clustering and graph methods for selecting tag SNPs. *Bioinformatics* **21(8)** 1735–1736.

[3] Avi-Itzhak, H.I., Su, X. and De La Vega, F.M. (2003). Selection of minimum subsets of single nucleotide polymorphisms to capture haplotype block diversity. *Pac Symp Biocomput.* 466–477.

[4] Azevedo, L., Suriano, G., van Asch, B., Harding, R. M. and Amorim, A. (2006). Epistatic interactions: how strong in disease and evolution? *Trends Genet.* **11** 585–598.

[5] Baker, S. G. (2005). A simple loglinear model for haplotype effects in a case-control study involving two unphased genotypes. *Statistical Applications in Genetics and Molecular Biology* **4(1)** 14. MR2138219

[6] Becker, T., Cichon, S., Jonson, E. and Knapp, M. (2005). Multiple testing in the context of haplotype analysis revisited: application to case-control data. *Annals of Human Genetics* **69** 747–756.

[7] Becker, T. and Knapp, M. (2004). A powerful strategy to account for multiple testing in the context of haplotype analysis. *Am J Hum Genet.* **75(4)** 561–570.

[8] Beckmann, L., Thomas, D.C., Fischer, C. and Chang-Claude, J. (2005). Haplotype sharing analysis using Mantel statistics. *Human Heredity* **59** 67–78.

[9] Benjamin, D. H. and Nicola, J. C. (2004). Principal component analysis for selection of optimal SNP-sets that capture intragenic genetic variation. *Genetic Epidemiology* **26(1)** 11–21.

[10] Bo, T. and Jonassen, I. (2002). New feature subset selection procedures for classification of expression profiles. *Genome Biology* **3(4)** research0017.

[11] Breiman, L. (2001). Random Forests. *Machine Learning* **45** 5–32.

[12] Breiman, L., Friedman, J. H., Olshen, R. A. and Stone, C. J. (1984). *Classification and Regression Tress* Wadsworth, Belmont.

[13] Brookes, A.J. (1999). Review: The essence of SNPs. *Gene* **234** 177–186.

[14] Burkett, K., McNeney, B. and Graham,J. (2004). A note on inference of trait associations with SNP haplotypes and other attributes in generalized linear models. *Human Heredity* **57** 200–206.

[15] Burton, P. R., Tobin, M.D. and Hopper, J.L. (2005). Key concepts in genetic epidemiology. *Lacent* **366** 941–951.

[16] Cardon, L. R. and Bell, J. I. (2001). Association study designs for complex diseases. *Nat Rev Genet* **2** 91–99.

[17] Carlson, C.S., Eberle, M.A., Rieder, M.J., Yi, Q., Kruglyak, L. and Nickerson D.A. (2004). Selecting a maximally informative set of single-nucleotide polymorphisms for association analyses using linkage disequilibrium. *Am J Hum Genet.* **74** 106–120.

[18] Chapman, J. M., Cooper, J. D., Todd, J. A. and Clayton, D. G. (2003). Detecting disease associations due to linkage disequilibrium using haplotype tags: a class of tests and the determinants of statistical power. *Hum. Hered.* **56** 18–31.

[19] Chatterjee, N., Kalaylioglu, Z., Moslehi, R., Peters, U. and

WACHOLDER, S. (2006). Powerful multilocus tests of genetic association in the presence of gene-gene and gene-environment interactions. *American Journal of Human Genetics* **79(6)** 1002–1016.

[20] CHENG, R., MA, J., ELSTON, R.C. and LI, M.D. (2005). Fine mapping functional sites or regions from case-Control data using haplotypes of multiple linked SNPs. *Annals of Human Genetics* **69(1)** 102–112.

[21] CLARK, T. G., DE IORIO, M., GRIFFITHS, R. C. and FARRALL, M. (2005). Finding associations in dense genetic maps: a genetic algorithm approach. *Human Heredity* **60** 97–108.

[22] COFFEY, C.S., HEBERT, P.R., RITCHIE, M.D., KRUMHOLZ, H.M., MORGAN, T.M., GAZIANO, J.M. RIDKER, P.M. and MOORE, J.H. (2004). An application of conditional logistic regression and multifactor dimensionality reduction for detecting gene-gene interactions on risk of myocardial infarction: The importance of model validation. *BMC Bioinformatics* **5** 49.

[23] CONNEELY, K. N. and BOEHNKE, M. (2005). Combining correlated p-values in trait-SNP association studies. *The American Society of Human Genetics 55th Annual Meeting, Salt Lake City, Utah* 184–189.

[24] CORES, C. and VAPNIK, V. N. (1995). Support Vector Networks. *Machine Learning* **20** 273–297.

[25] DALY, M. J., RIOUX, J. D., SCHAFFNER, S. F., HUDSON, T. J. and LANDER, E.S. (2001). High-resolution haplotype structure in the human genome. *Nat. Genet.* **29** 229–232.

[26] DEMBO, A. and KARLIN, S. (1992). Poisson approximations for r-scan processes. *The Annals of Applied Probability* **2** 329–357. MR1161058

[27] DUDBRIDGE, F. and KOELEMAN, B. P. C. (2004). Efficient computation of significance levels for multiple associations in large studies of correlated data, including genomewide association studies. *American Journal of Human Genetics* **75(3)** 424–435.

[28] DURRANT, C., ZONDERVAN, K. T., CARDON, L. R., HUNT, S., DELOUKAS, P. and MORRIS, A. P. (2004). Linkage Disequilibrium Mapping via Cladistic Analysis of Single-Nucleotide Polymorphism Haplotypes. *Am. J. Hum. Genet.* **75** 35–43.

[29] FU, R., DEY, D. K. and HOLSINGER, K. E. (2005). Bayesian models for the analysis of genetic structure when populations are correlated. *Bioinformatics* **21(8)** 1516–1529.

[30] GOPALAKRISHNAN, S. and QIN, Z. S. (2006). TagSNP Selection Based on Pairwise LD Criterion and Power Analysis in Association Studies *Pacific Sym. Biocomputing* **11** 511–522.

[31] GREENSPAN, G. and GEIGER, D. (2004). Model-based inference of haplotype block variation. *J. Comp. Biol.* **11** 493–504.

[32] GREENSPAN, G. and GEIGER, D. (2006). Modeling Haplotype Block Variation Using Markov Chains. *Genetics* **172(4)** 2583–2599.

[33] GUYON, I., WESTON, J., BARNHILL, S. and VAPNIK, V. N. (2002). Gene Selection for Cancer Classification using Support Vector Machines. *Machine Learning* **46(1–3)** 389–422.

[34] HALLDORSSON, B. V., BAFNA, V., LIPPERT, R., SCHWARTZ, R., DE LA VEGA, F. M., CLARK, A. G. and ISTRAIL, S. (2004). Optimal haplotype block-free selection of tagging SNPs for genomewide association studies. *Genome Res* **14** 1633–1640.

[35] HALPERIN, E., KIMMEL, G. and SHAMIR, R. (2005). Tag SNP Selection in Genotype Data for Maximizing SNP Prediction Accuracy. *Bioinformatics* **21(suppl 1)** 195–203.

[36] HAMPE, J., SCHREIBER, S. and KRAWCZAK, M. (2003). Entropy-based SNP selection for genetic association studies. *Hum Genet.* **114** 36–43.

[37] HE, J. and ZELIKOVSKY, A. (2006). MLR-tagging informative SNP selection for unphased genotypes based on multiple linear regression. *Bioinformatics* **22(20)** 2558–2561.

[38] HIRSCHHORN, J. N. and DALY, M. J. (2005). Genome-wide association studies for common diseases and complex traits. *Nature Reviews Genetics* **6** 95–108.

[39] HOH, J. and OTT, J. (2000). Scan statistics to scan markers for susceptibility genes. *Proc Nat Acad Sci* **97** 9615–9617.

[40] HOWIE, B. N., CARLSON, C. S., RIEDER, M. J. and NICKERSON, D. A. (2006). Efficient selection of tagging single-nucleotide polymorphisms in multiple populations. *Human Genetics* **120(1)** 58–68.

[41] HUBLEY, R. M., ZITZLER, E. and ROACH, J. C. (2003). Evolutionary algorithms for the selection of single nucleotide polymorphisms. *BMC Bioinformatics* **4** 30–39.

[42] HUNG, R. J., BRENNAN, P., MALAVEILLE, C., PORRU, S., DONATO, F., BOFFETTA, P. and WITTE, J. S. (2004). Using hierarchical modeling in genetic association studies with multiple markers: application to a case-control study of bladder cancer. *Cancer Epidemiology Biomarkers and Prevention* **13(6)** 1013–1021.

[43] HUNTER, D. J. (2005). Gene-environment interactions in human diseases. *Nature Reviews Genetics* **6** 287–298.

[44] INZA, I., SIERRA, B., BLANCO, R. and LARRANAGA, P. (2002). Gene selection by sequential search wrapper approaches in microarray cancer class prediction *Journal of Intelligent and Fuzzy Systems* **12(1)** 25–34.

[45] IOANNIDIS, J. P., GWINN, M., LITTLE, J., HIGGINS, J. P., BERNSTEIN, J. L., BOFFETTA, P., BONDY, M., BRAY, M. S., BRENCHLEY, P.E., BUFFLER, P. A. ET AL. (2006). Human Genome Epidemiology Network and the Network of Investigator Networks, A road map for efficient and reliable human genome epidemiology. *Nature Genetics* **38(1)** 3–5.

[46] JUDSON, R, SALISBURY, B., SCHNEIDER, J., WINDEMUTH, A. and STEPHENS, J. C. (2002). How many SNPs does a genome-wide haplotype map require? *Pharmacogenomics* **3** 379–391.

[47] KASABOV, N. (2002). *Evolving Connectionist Systems: Methods and Applications in Bioinformatics, Brain Study and Intelligent Machines.* London-New York, Springer-Verlag. MR2375896

[48] KE, X. and CARDON, L. R. (2003). Efficient selective screening of haplotype tag SNPs. *Bioinformatics* **19** 287–288.

[49] KNORR-HELD, L. and RUE, H. (2002). On block updating in Markov random field models for disease mapping. *Scandinavian Journal of Statistics* **29(4)** 597–614. MR1988414

[50] KRINA, T., ZONDERVAN, L. and CARDON, T. (2004). The complex interplay among factors that influence allelic association. *Nature Reviews Genetics* **5(2)** 89–100.

[51] KRISHNAPURAM, B. and CARIN, L. (2005). Sparse Multinomial Logistic Regression: Fast Algorithms and Generalization Bounds. *IEEE Transactions on Pattern Analysis and Machine Intelligence* **27(6)**.

[52] LAL, T. N., CHAPELLE, O., WESTON, J. and ELISSEEFF, A. (2006). Embedded methods. Feature Extraction: Foundations and Applications. In Guyon, I., Gunn, S., Nikravesh, M. Zadeh, L. A. (Eds.) Springer, Berlin, Germany.

[53] LAM, J. C., ROEDER, K. and DEVLIN, B. (2000). Haplotype fine mapping by evolutionary trees. *Am. J. Hum. Genet.* **66 (2)** 659–673.

[54] LEVIN, A. M., GHOSH, D., CHO, K. R. and KARDIAS. L. R. (2005). A model-based scan statistics for identifying extreme chromosomal regions of gene expression in human tumors. *Bioinformatics* **21** 2867–2874.

[55] LI, J. and JIANG, T. (2005). Haplotype-based linkage disequilibrium mapping via direct data mining *Bioinformatics* **21** 4384–4393.

[56] LIANG, Y. and KELEMEN, A. (2005). Temporal Gene Expression Classification with Regularised Neural Network. *International Journal of Bioinformatics Research and Applications* **1(4)** 399–413.

[57] LIN, Z. and ALTMAN, R. B. (2004). Finding haplotype tagging SNPs by use of principal components analysis. *Am. J. Hum. Genet.* **75** 850–861.

[58] LIU, J. S., SABATTI, C., TENG, J., KEATS, B. J. and RISCH, N. (2001). Bayesian analysis of haplotypes for linkage disequilibrium mapping. *Genome Research* **11 (10)** 1716–1724.

[59] LIU, Z. and LIN, S. (2005). Multilocus LD measure and tagging SNP selection with generalized mutual information. *Genet Epidemiol.* **29** 353–364.

[60] LONG, A., MANGALAM, H., CHAN, B., TOLLERI, L., HATFIELD, G. and BALDI, P. (2001). Improved statistical inference from DNA microarray data using analysis of variance and a Bayesian statistical framework. *J. Biol. Chem.* **276** 19937–19944.

[61] MANNILA, H., KOIVISTO, M., PEROLA, M., VARILO, T., HENNAH, W., EKELUND, J., LUKK, M., PELTONEN, L. and UKKONEN, E.(2003). Minimum description length block finder, a method to identify haplotype blocks and to compare the strength of block boundaries. *Am. J. Hum. Genet.* **73** 86–94.

[62] MOLITOR, J., MARJORAM, P. and THOMAS, D. (2003). Fine-Scale Mapping of Disease Genes with Multiple Mutations via Spatial Clustering Techniques. *Am. J. Hum. Genet.* **73** 1368–1384.

[63] MONARI, G. and DREYFUS, G. (2000). Withdrawing an example from the training set: an analytic estimation of its effect on a nonlinear parameterized model. *Neurocomputing Letters* **35** 195–201.

[64] MOORE, J. H. (2007). Genome-wide analysis of epistasis using multifactor dimensionality reduction: feature selection and construction in the domain of human genetics. In: Zhu, Davidson (eds.) Knowledge Discovery and Data Mining: Challenges and Realities with Real World Data, IGI, (in press).

[65] MOORE, J. H. and WHITE, B. C. (2006). Exploiting expert knowledge for genome-wide genetic analysis using genetic programming. In: Runarsson et al. (eds.) Parallel Problem Solving from Nature - PPSN IX, Lecture Notes in Computer Science 4193, 969–977.

[66] MOORE, J. H. and WILLIAMS, S. M. (2002). New strategies for identifying gene-gene interactions in hypertension. *Ann Med.*

[67] MOTSINGER, A. A., LEE, S. L., MELLICK, G. and RITCHIE, M. D. (2006). PNN: Power studies and applications of a neural network method for detecting gene-gene interactions in studies of human disease. *BMC Bioinformatics* **7(1)** 39–50.

[68] NEALE, B. and SHAM, P. (2004). The future of association studies: Gene-based analysis and replication. *American Journal of Human Genetics* **75** 353–362.

[69] NEWTON, M. A., KENDZIORSKI, C. M., RICHMOND, C. S., BLATTNER, F. R. and TSUI, K. W. (2001). On differential variability of expression ratios: improving statistical inference about gene expression changes from microarray data. *Journal of Computational Biology* **8(1)** 37–52.

[70] NYHOLT, D. R. (2004). A simple correction for multiple testing for single-nucleotide polymorphisms in linkage disequilibrium with each other. *American Journal of Human Genetics* **74(4)** 765–769.

[71] OTT, J. (2001). Neural networks and disease association studies. *merican Journal of Medical Genetics* **105 (1)** 60–61.

[72] PARK, M. and HASTIE, T. (2006). Regularization Path Algorithms for Detecting Gene Interactions, preprint.

[73] PAVLIDIS, P. and NOBLE, W. S. (2001). Analysis of strain and regional variation in gene expression in mouse brain. *Genome Biology* **2(10)** research0042.1-0042.15.

[74] PEDRYCZ, W. (1997). *Computational Intelligence: An Introduction.* Boca Raton, FL, CRC.

[75] RISCH, N. J. (2000). Searching for genetic determinants in the new millennium. *Nature* **405** 847–856.

[76] RISCH, N. and MERIKANGAS, K. (1996). The future of genetics studies of complex human diseases. *Science* **273** 1516–1517.

[77] RITCHIE, M. D., HAHN, L. W. and MOORE, J. H. (2003a). Power of multifactor dimensionality reduction for detecting gene-gene interactions in the presence of genotyping error, missing data, phenocopy, and genetic heterogeneity. *Genet Epidemiol.* **24** 150–157.

[78] RITCHIE, M. D., WHITE, B. C., PARKER, J. S., HAHN, L. W. and MOORE, J. H. (2003b). Optimization of neural network architecture using genetic programming improves detection and modeling of gene-gene interactions in studies of human diseases. *BMC Bioinformatics* **4** 28–38.

[79] RIVALS, I. and PERSONNAZ, L. (2003). MLPs (Mono-Layer Polynomials and Multi-Layer Perceptrons) for Nonlinear Modeling. *Journal of Machine Learning Research* **3** 1383–1398.

[80] SALYAKINA, D., SEAMAN, S. R., BROWNING, B. L., DUDBRIDGE, F. and MULLER-MYHSOK, B. (2005). Evaluation of Nyholt's procedure for multiple testing correction. *Human Heredity* **60(1)** 19–25.

[81] SCHAID, D. J. (1996). General score tests for associations of genetic markers with disease using cases and their parents. *Genetic Epidemiology* **13** 423–449.

[82] SCHAID, D. J., ROWLAND, C. M., TINES, D. E., JACOBSON, R. M. and POLAND, G. A. (2002). Score test for association between traits and haplotypes when linkage phase is ambiguous. *Am J Hum Genet* **70** 425–439.

[83] SCHWENDER, H. and ICKSTADT, K. (2006). Identification of SNP Interactions Using Logic Regression, http://www.sfb475.uni-dortmund.de/berichte/tr31-06.pdf, accessed on Oct.-31-2006.

[84] SEAMAN, S.R. and MULLER-MYHSOK, B. (2005). Rapid simulation of P values for product methods and multiple-testing adjustment in association studies. *American Journal of Human Genetics* **76** 399–408.

[85] SEBASTIANI, P., LAZARUS, R., WEISS, S. T., LUNKEL, L. M., KOHANE, I. S. and ROMANI, M. F. (2003). Minimal haplotype tagging. *Proc. Natl. Acad. Sci. USA* **100** 9900–9905.

[86] SHRIVER, M., MEI, R., PARRA, E. J., ET AL., (2005). Large-scale SNP analysis reveals clustered and continuous patterns of human genetic variation. *Human Genomics* **2(2)** 81–89.

[87] SONG, K. and ELSTON, R. C. (2006). A powerful method of combining measures of association and Hardy-Weinberg disequilibrium for fine-mapping in case-control studies. *Statistics in Medicine* **25(1)** 105–126. MR2222077

[88] STEPHENS, M. and DONNELLY, P. (2000). Inference in molecular population genetics. *J R Stat Soc B* **62** 605–655. MR1796282

[89] STRAM, D. O., HAIMAN, C. A., HIRSCHHORN, J. N., ALTSHULER, D., KOLONEL, L. N., HENDERSON, B. E. and PIKE, M. C. (2003). Choosing haplotype-tagging SNPs based on unphased genotype data using preliminary sample of unrelated subjects with an example from the multiethnic cohort study. *Hum. Hered.* **55** 27–36.

[90] SUN, W. and CAI, T. (2007). Oracle and adaptive compound decision rules for false discovery rate control. *J. American Statistical Association* **102** 901–912.

[91] SUN, Y., LEVIN, A., BOERWINKLE, E., ROBERTSON, H. and KARDIA, S. (2006). A scan statistic for identifying chromosomal patterns of SNP association. *Genetic Epidemiology* **30** 627–635.

[92] TAN, P., STEINBACH, M. and KUMAR, V. (2005). Introduction to Data Mining, Addison-Wesley, pp. 76–79.

[93] THE INTERNATIONAL HAPMAP CONSORTIUM (2005). A haplotype map of the human genome. *Nature* **437** 1299–1320.

[94] THE INTERNATIONAL HAPMAP CONSORTIUM (2004). Integrating ethics and science in the International HapMap Project. *Nat Rev Genet* **5** 467–475.

[95] THE INTERNATIONAL HAPMAP CONSORTIUM (2003). The International HapMap Project. *Nature* **426** 789–796.

[96] THOMAS, D. C., STRAM, D. O., CONTI, D., MOLITOR, J. and MARJORAM, P. (2003). Bayesian spatial modeling of haplotype associations. *Human Heredity* **56** 32–40.

[97] TIBSHIRANI, R. (1996). Regression shrinkage and selection via the lasso. *J. Royal. Statist. Soc B.* **58(1)** 267–288. MR1379242

[98] TIBSHIRANI, R. (1997). The lasso method for variable selection in the Cox model. *Statistics in Medicine* **16** 385–395.

[99] TOIVONEN, H. T., ONKAMO, P., VASKO, K., OLLIKAINEN, V., SEVON, P., MANNILA, H., HERR, M. and KERE, J. (2000). Data mining applied to linkage disequilibrium mapping. *Am. J. Hum. Genet.* **67(1)** 133–145.

[100] TZENG, J. N., WANG, C. H., KAO, J. T. and HSIAO, C. K. (2006). Regression-based association analysis with clustered haplotypes through use of genotypes. *American Journal of Human Genetics* **78(2)** 231–242.

[101] VAPNIK, V. N. (1995). The Nature of Statistical Learning Theory. Springer-Verlag, New York MR1367965

[102] VAPNIK, V. N. (1998). Statistical Learning Theory. Wiley, New York. MR1641250

[103] VERZILLI, C. J., STALLARD, N. and WHITTAKER, J. C. (2006). Bayesian graphical models for genomewide association studies. *American Journal of Human Genetics* **79(1)** 100–112.

[104] WALLENSTEIN, S. and NEFF, N. (1987). An approximation for the distribution of the scan statistic. *Stat Med* **6** 197–207.

[105] WANG, L., ZHU, J. and ZOU, H. (2006). Doubly regularized support vector machine. *Statistica Sinica* **16** 589–615. MR2267251

[106] WESSEL, J. and SCHORK, N. J. (2006). Generalized Genomic Distance Based Regression Methodology for Multilocus Association Analysis. *American Journal of Human Genetics* **79(5)** 792–806.

[107] WESTON, J., MUKHERJEE, S., CHAPELLE, O., PONTIL, M., POGGIO, T. and VAPNIK, V. (2000). Feature Selection for SVMs. In S. A. Solla, T. K. Leen, and K. R. Muller, (eds), Advances in Neural Information Processing Systems, volume 12, 526–532, Cambridge, MA, USA. MIT Press.

[108] WITTE, J. S. and FIJAL, B. A. (2001). Introduction: Analysis of Sequence Data and Population Structure. *Genetic Epidemiology* **21** 600–601.

[109] YU, J. and CHEN, X. W. (2005). Bayesian Neural Network Approaches to Ovarian Cancer Identification from High-resolution Mass Spectrometry Data. *Bioinformatics* **21 (suppl-1)** i487–i494.

[110] YUAN, M. and LIN, Y. (2006). Model selection and estimation in regression with grouped variables. *Journal of the Royal Statistical Society: Series B (Statistical Methodology)* **68 (1)** 49–67. MR2212574

[111] ZAYKIN, D. V., WESTFALL, P. H., YOUNG, S. S., KARNOUB, M. A., WAGNER, M. J. and EHM, M. G. (2002b). Testing Association of Statisti-

cally Inferred Haplotypes with Discrete and Continuous Traits in Samples of Unrelated Individuals. *Hum Hered* **53** 79–91.

[112] ZAYKIN, D. V. and ZHIVOTOVSKY, L. A. (2005). Ranks of genuine associations in whole-genome scans. *Genet* **171** 813–823.

[113] ZAYKIN, D. V., ZHIVOTOVSKY, L. A., ET AL. (2002a). Truncated product method for combining P-values. *Genet Epidemiol* **22** 170–185.

[114] ZHANG, K. and JIN, L. (2003). HaploBlockFinder: Haplotype block analysis. *Bioinformatics* **19** 1300–1301.

[115] ZHANG, K., QIN, Z., LIU, J., CHEN, T., WATERMAN, M. S. and SUN, F. (2004). Haplotype Block Partitioning and Tag SNP Selection Using Genotype Data and Their Applications to Association Studies. *Genome Res.* **14** 908–916.

[116] ZHANG, Y., NIU, T. and LIU, J. (2006). A coalescence-guided hierarchical Bayesian method for haplotype inference. *American Journal of Human Genetics* **79(2)** 313–322.

[117] ZHAO, J., BOERWINKLE, E. and XIONG, M. (2005). An entropy-based statistic for genomewide association studies. *American Journal of Human Genetics* **77** 27–40.